\documentstyle[sprocl,epsf]{article}
\def\be{\begin{equation}}
\def\ee{\end{equation}}
\def\bea{\begin{eqnarray}}
\def\eea{\end{eqnarray}}
\newcommand{\pom}{I\!\! P}
\input{psfig}
\begin{document}
\title{RESULTS ON DIFFRACTION
\footnote{Presented at the XVII$^{th}$ International Conference on Physics in 
Collision, Bristol, UK, 25-27 June 1997.}
}
\author{K. GOULIANOS}
\address{The Rockefeller University, 1230 York Avenue\\ 
New York, NY 10021, USA}
\maketitle
\abstracts{Results on hard diffraction from HERA and 
$p\bar p$ Colliders are reviewed
with emphasis on the factorization properties of the diffractive structure
function of the proton and on the structure of the pomeron.
}
\section{Introduction}
In the past few years the field of diffraction has experienced a renaissance.
Experiments at $p\bar p$ Colliders and at HERA have been probing 
the diffractive structure function of the proton,  
shedding light at the intricate interplay 
between soft and hard diffraction. 
While a firm QCD based theoretical interpretation 
of the experimental results 
is still lacking, a rather clear phenomenological picture of the connection 
between soft and hard diffraction  
has emerged. In this picture, the (virtual) pomeron, which is presumed to
be exchanged in diffractive processes, appears as a simple 
color-singlet construct of quarks and gluons. 
Its partonic structure is surprisingly hard 
(each parton carries a large fraction of the pomeron momentum) and, 
unlike that of real hadrons, persists being hard at high values of $Q^2$.  
The wealth of the accumulated experimental data allows 
questions to be asked about the factorization properties of the 
diffractive structure function of the proton, the uniqueness of the 
pomeron structure, and the momentum sum rule for the pomeron. 
In this paper, we review the experimental results
with emphasis on what is deemed to be directly relevant to 
answering these questions. 
After a brief general discussion  of diffraction, we present results 
from the $S\bar ppS$ Collider,  from HERA and 
from the Tevatron, relate the HERA and Collider 
results, and draw conclusions.

\section{Soft and Hard Diffraction}
Single diffraction (SD) dissociation accounts for approximately 10\% 
of the total hadronic cross sections~\cite{KG}. At high energies, 
the cross section for SD is 
dominated by pomeron ($\pom$) exchange. For $p\bar p\rightarrow pX$, it has the 
form
\begin{equation}
\frac{d^2\sigma_{SD}}{d\xi dt}=
\frac{{\beta_{\pom p}^2(t)}}{16\pi}\;\xi^{1-2\alpha(t)}
\left[ \beta_{\pom p}(0)\,g(t)\cdot
({s'}/{s_0})^{\alpha(0)-1}\right]
\label{diffractive}
\end{equation}
where $\alpha(t)=1+\epsilon+\alpha' t$ is the pomeron Regge trajectory, 
$\beta_{\pom p}(t)$ is the  coupling of the pomeron to the proton,
$g(t)$ the triple-pomeron coupling,
$s'$ the $\pom-\bar p$ center of mass energy squared,
$\xi\equiv x_{\pom}=s'/s\approx M^2/s$ the fraction of 
the momentum of the proton carried
by the pomeron,
$M$ the diffractive mass and $s_0$ an energy scale not specified by the 
theory. Experimentally, 
the triple-pomeron coupling was found not to depend on $t$~\cite{KG}.
A recent determination of $\epsilon$ from a global fit to 
$p^{\pm}p,~\pi^{\pm}p$ and $K^{\pm}p$ total cross sections and $\rho$-values 
yielded~\cite{CMG} $\epsilon=0.104\pm 0.002$; from elastic scattering data, 
$\alpha'\approx 0.25$ GeV$^{-2}$. 
\vglue -0.1in
\begin{figure}[h,t]
\centerline{\psfig{figure=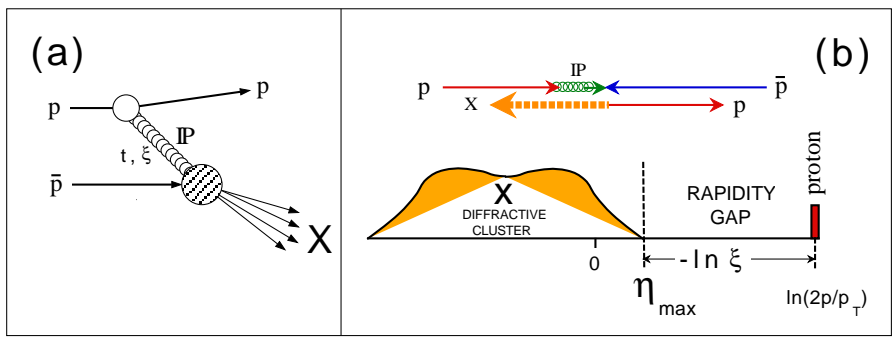,width=5in}}
\vspace*{-4.75in}
\caption{(a) Feynman diagram and (b) event topology for $p\bar p$ 
diffraction dissociation.}
\label{sdd_rapgap}
\end{figure}
The term in brackets in 
(\ref{diffractive})
is identified as the $\pom-\bar p$ total cross section 
(see Fig.~\ref{sdd_rapgap}a) and therefore the factor 
\begin{equation}
f_{\pom /p}(\xi,t)\equiv \frac{{\beta_{\pom p}^2(t)}}{16\pi}\;\xi^{1-2\alpha(t)}
=\frac{{\beta_{\pom p}^2(0)}}{16\pi}\;\xi^{1-2\alpha(t)}F^2(t)
=K\;\xi^{1-2\alpha(t)}F^2(t)
\label{flux}
\end{equation}
is interpreted as a ``pomeron flux" per proton. In the 
Ingelman-Schlein (IS) model~\cite{IS} the pomeron flux is used as 
a luminosity factor in calculating the rates of hard processes 
in diffraction dissociation (hard diffraction). Such calculations are 
usually performed using the Monte Carlo simulation program POMPYT~\cite{POMPYT},
which combines routines that generate the pomeron $\xi$ and $t$ 
with the program PYTHIA~\cite{PYTHIA}, which ``performs" the 
collision.

At small $t$, $F^2(t)\approx e^{\textstyle 4.6t}$.
Donnachie and Landshoff (DL) proposed a model~\cite{DLF} 
in which $F(t)$ is identified as the isoscalar form factor 
measured in electron-nucleon 
scattering, 
\begin{equation}
F_1(t)=\frac{4m^2-2.8t}{4m^2-t}\left[\frac{1}{1-t/0.7}\right]^2
\label{F1}
\end{equation}
where $m$ is the mass of the proton. In the DL model, 
the factor $K$ in the pomeron flux is 
$K_{DL}={(3\beta_{\pom q})^2}/{4\pi^2}$,
where $\beta_{\pom q}$ is the $\pom$-quark coupling.

Figure~\ref{sdd_rapgap}b shows
the event topology in pseudorapidity space for $p\bar p\rightarrow pX$.
It is characterized by a ``leading" proton at large
rapidity, and a rapidity gap (absense of particles) in the region 
between the leading proton
and the ``diffractive cluster" of particles resulting from the $\pom-\bar p$
collision. These characteristics are used to ``tag" diffractive events.

The normalization of the pomeron flux depends on $\beta_{\pom p}(0)$, 
which is determined from the experimentally measured $pp$ total cross section,
$\sigma_T^{pp}=\beta^2_{\pom p}(0)\cdot (s/s_0)^{\epsilon}$. 
Therefore, $\beta_{\pom p}(0)$ 
depends on the value of the energy scale $s_0$, which, 
as mentioned above, is not given by 
the theory ($s_0$ is usually 
set to 1 GeV$^2$, the hadron mass scale, but this is only {\em a convention}).
Thus, the normalization of the pomeron flux is  
unknown, and hence only predictions for {\em relative} hard diffraction 
rates are possible in the IS model.

The flux normalization uncertainty is resolved in the flux renormalization 
model of Goulianos~\cite{R}. The {\em renormalized} flux is defined as 
\begin{equation}
f_N(\xi,t)\equiv D_N\cdot f_{\pom/p}(\xi,t;s_0);\;\;\;\;
D_N=1/\displaystyle{\int_{\xi_{min}}^{0.1}}\displaystyle{\int_0^{\infty}}
f_{\pom/p}(\xi,t;s_0)d\xi dt
\label{fluxN}
\end{equation}
where $ f_{\pom/p}(\xi,t;s_0)$ is the {\em standard} flux and 
$D_N$ is the renormalization or flux discrepancy factor~\cite{R}.
The integration over $\xi$ is carried out between the 
lowest kinematically allowed 
value, $\xi_{min}=M_0^2/s$, where $M_0^2=1.5$ GeV$^2$ is the 
effective diffractive threshold~\cite{R},  and $\xi_{max}=0.1$, which is 
the ``coherence limit"~\cite{KG}. 
Such a normalization, which 
corresponds to {\em at most} one pomeron per proton, leads to interpreting the 
pomeron flux as a probability density describing the $\xi$ and $t$ 
distributions of the exchanged pomeron. 

As indicated explicitly in (\ref{fluxN}), the renormalized flux 
does not depend on the energy scale $s_0$.  
Thus, using this flux, {\em robust} predictions for hard 
diffraction can be obtained not only for relative but also for 
absolute rates. 

The renormalized pomeron flux model is supported by the $s$-dependence 
of the total and differential $pp/\bar pp$ SD cross sections.
With the standard flux, the total and $t=0$ differential 
diffractive cross sections vary as 
\begin{eqnarray}
\sigma_{SD}(s)\sim s^{2\epsilon}
&\frac{d^2\sigma_{SD}}{dtdM^2}|_{t=0}\sim \frac{s^{2\epsilon}}
{(M^2)^{1+\epsilon}}
\end{eqnarray}
The $\sim s^{2\epsilon}$ dependence 
eventually leads to diffractive cross sections larger than the total and 
therefore to violation of unitarity. The flux renormalization factor
varies as $D_N\sim s^{-2\epsilon}$, canceling the $s^{2\epsilon}$ growth 
of the standard flux and preserving unitarity.
Figure~\ref{sdt_m2} shows that the standard flux fails to describe the 
data, but the renormalized flux predicts the 
observed $s$-dependence~\cite{R,KGSX,KGDIS}.

In terms of the rapidity gap, $\Delta y$, the pomeron flux is given by
\begin{eqnarray}
\Delta y=\ln\frac{1}{\xi}
&\Rightarrow
&f_{\pom/p}(\Delta y,t)=K\cdot e^{\textstyle 2(\epsilon+\alpha't)\Delta y}
\cdot F^2(t)
\label{fluxY}
\end{eqnarray}
Since $\epsilon$ is small and $\alpha'\overline{|t|}\ll \epsilon$, the flux 
varies slowly with $\Delta y$, approximately as $1+2\epsilon\Delta y$.
In this ``gap representation", the renormalized flux will be referred to 
as the {\em scaled gap probability}.

In the following sections, both the standard and renormalized flux 
predictions for hard diffraction rates will be compared with data. \\
{\hspace*{-0.3in}\begin{minipage}[t]{3.75in}
.\\
\vglue -0.7truein
{\psfig{figure=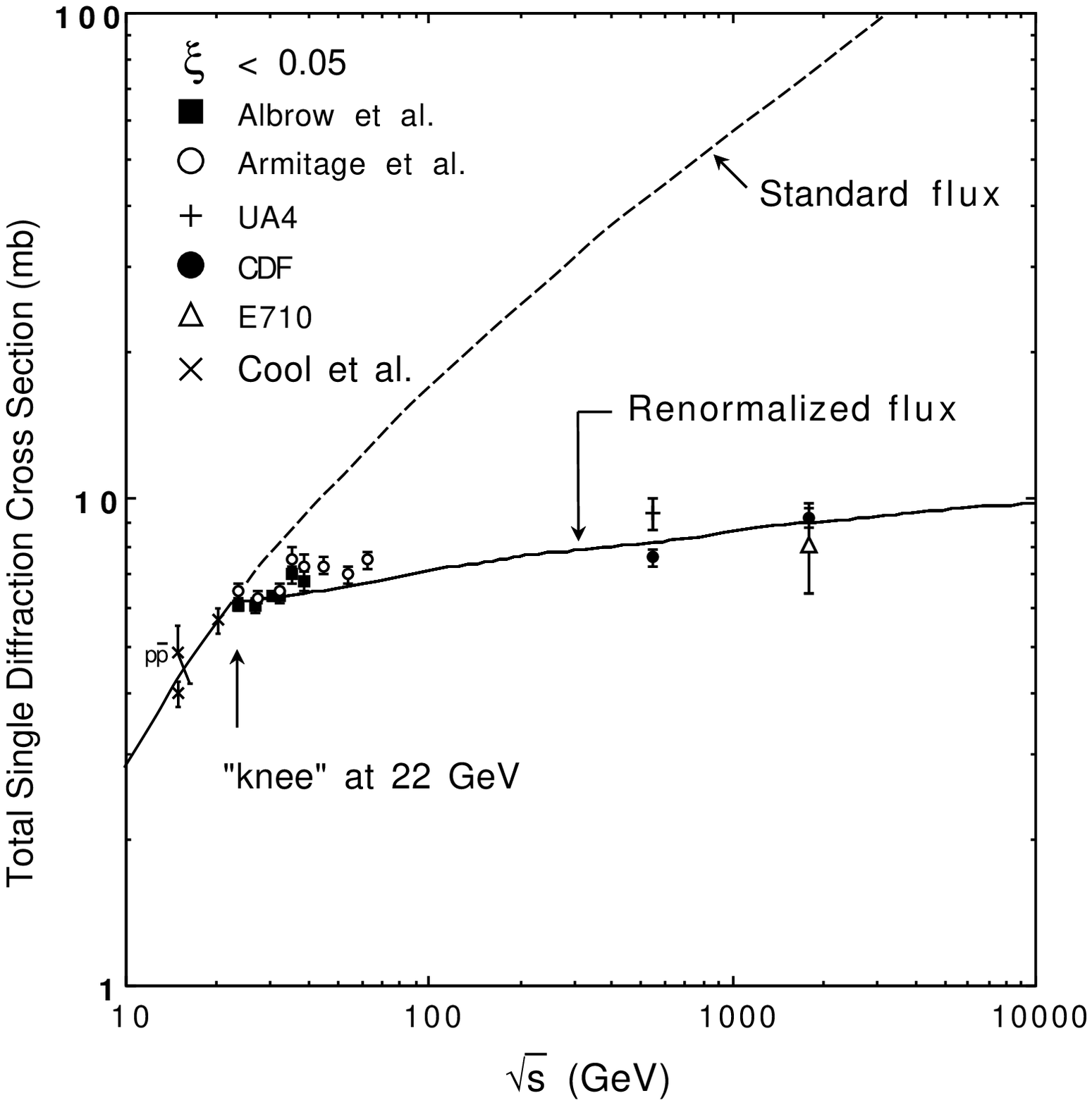,height=3.6in,width=3.0in}}
\end{minipage}
\   \
{\hspace{-1.1in}\begin{minipage}[t]{2.65in}
.\\
{\psfig{figure=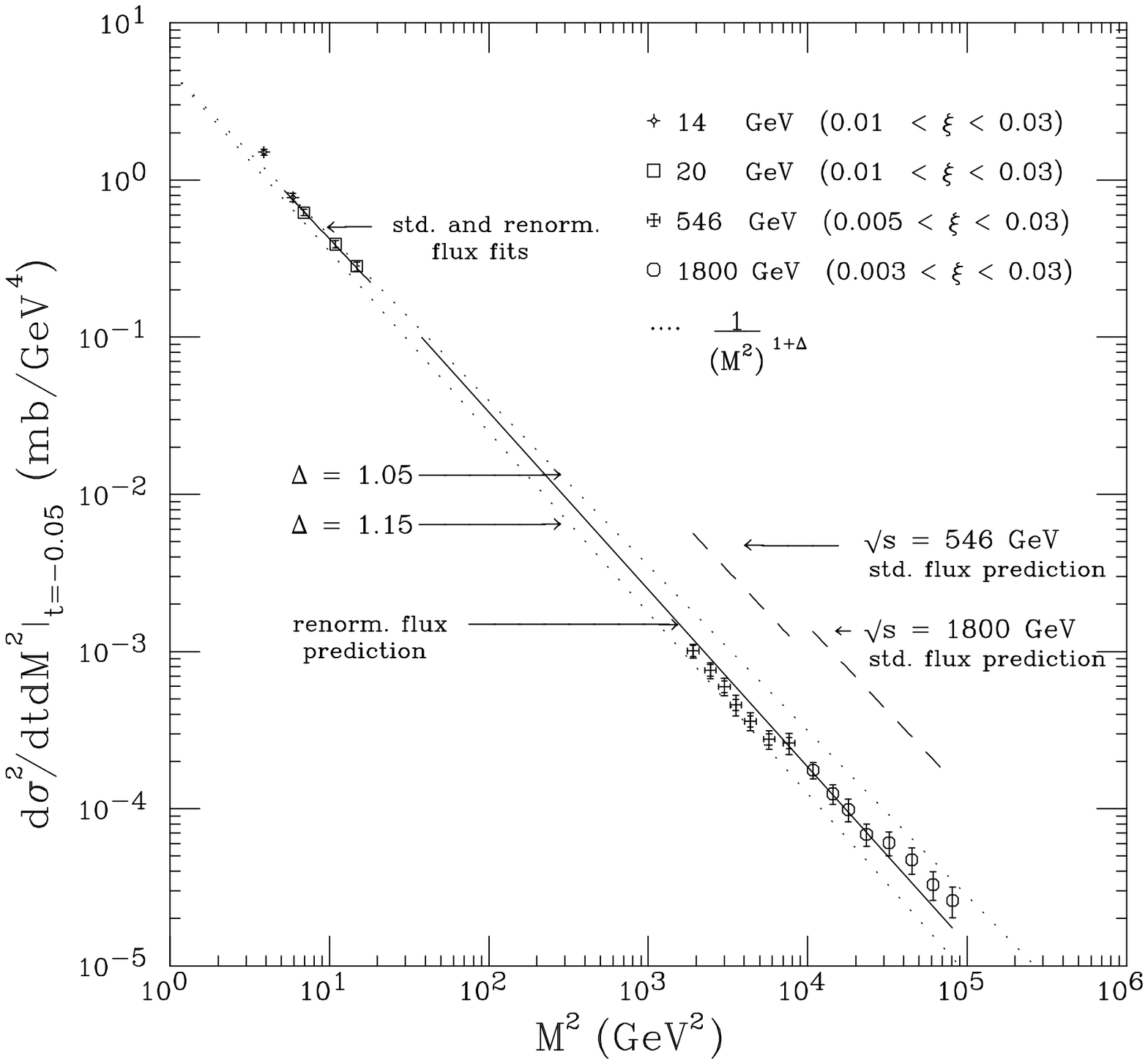,height=2.4in}}
\end{minipage}}}
\vglue -0.8in

\vglue -0.25in
\begin{figure}[h,t]
\caption{$p-p(\bar p)$ diffractive cross sections: {\em (left)} 
$2\sigma_{SD}(s)$; {\em (right)} $d^2\sigma/dtdM^2$ at t=-0.05.}
\label{sdt_m2}
\end{figure}
\section{Results from the $S\bar ppS$ Collider}
At the CERN $S\bar ppS$ Collider~\footnote{This section is an 
excerpt from Ref.~\cite{pbp2}}, 
the UA8 experiment 
pioneered hard diffraction studies
by observing high-$p_T$ jet production in the
process $p+\bar{p}\rightarrow p+Jet_1+Jet_2+X$ 
at $\sqrt{s}=630$ GeV.  Events with two jets of
$p_T>8$ GeV were detected
in coincidence with a high-$x_F$ proton,  whose momentum and angle
were measured in a forward ``roman pot" spectrometer. The event sample spanned
the kinematic range
$0.9<x_p<0.94$ and $0.9<|t|<2.3\;\mbox{GeV}^2$.
By comparing the $x_F$ distribution of the sum of the jet
momenta in the pomeron-proton rest frame with Monte Carlo
distributions generated with
different pomeron structure functions, UA8 concluded \cite{UA8}
that the partonic
structure of the pomeron is $\sim 57$\% {\em hard} [$6\beta (1-\beta)$],
$\sim 30$\% {\em superhard} [$\delta (1-\beta)$],
and $\sim 13$\% {\em soft} [$6(1-\beta)^5$], where $\beta$ is the fraction 
of the momentum of the pomeron carried by its parton constituent.
However, 
the measured dijet production rate was found to be smaller than 
the rate predicted by POMPYT, using the standard flux and assuming 
a hard-quark(gluon) pomeron, 
by a (discrepancy) factor of~\cite{UA8-EPS} $0.46\pm0.08\pm0.24$  
($0.19\pm0.03\pm0.10$). Using the renormalized pomeron flux, the 
discrepancy factor becomes consistent with unity~\cite{R}.
\section{Results from HERA}
At HERA, where $\sim 28$ GeV electrons are brought into collision with
$\sim 800$ GeV protons ($\sqrt{s}\approx 300$ GeV),
diffraction has been studied both in photoproduction and
in high $Q^2$ deep inelastic scattering (DIS). The H1 and ZEUS 
Collaborations have measured the diffractive structure function of the 
proton and its 
internal factorization properties. Below, we review the results of 
these measurements
and their relevance to the structure of the pomeron.
\subsection{Diffractive DIS kinematics}
\begin{minipage}[t]{2.8in}
\vglue -0.625in
\hspace*{-0.37in}\psfig{figure=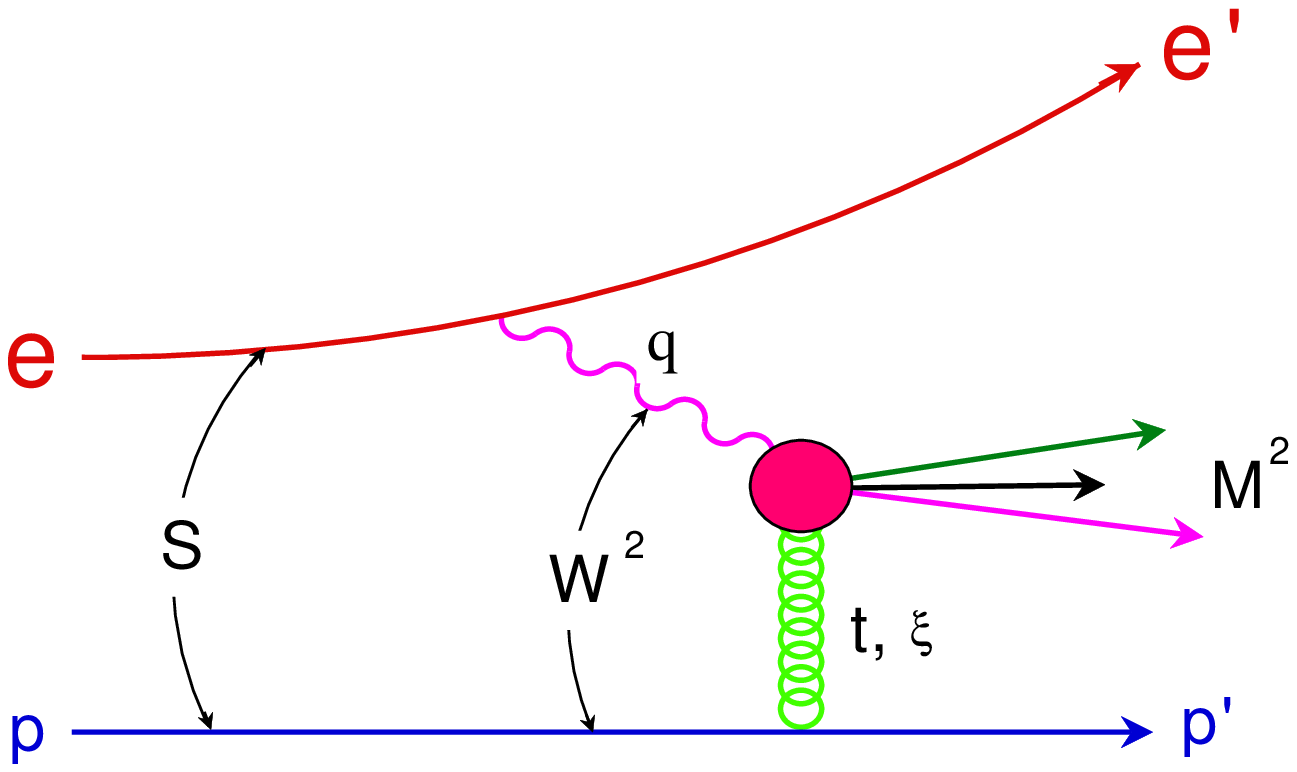,width=2.5in}
\vglue -7.375in
\end{minipage}
\  \
\begin{minipage}[t]{1.75in}
Figure~3: Schematic diagram of a diffractive 
DIS collision involving a virtual photon,
emitted by an electron, and a virtual pomeron,
emitted by a proton.
\end{minipage}
\vglue -1.5in

In addition to the standard DIS kinematical variables (see Fig.~3),
\begin{equation}
Q^2=-q^2\;\;\;\;x_{bj}=\frac{Q^2}{\textstyle 2\,q\cdot p}
\;\;\;\;y=\frac{\textstyle q\cdot p}
{\textstyle e\cdot p}
\;\;\;\;W^2=(q+p)^2\;\;\;\;s=(e+p)^2
\end{equation} 
the following variables are used to describe {\em diffractive} DIS:
\begin{equation}
\beta=\frac{Q^2}{2\,q\cdot(p-p')}=\frac{Q^2}{M_X^2+Q^2-t}
\;\;\;\;\xi=\frac{q\cdot (p-p')}{q\cdot p}
=\frac{M_X^2+Q^2-t}{W^2+Q^2-m_p^2}
\label{DISD}
\end{equation}
Note that $x_{bj}=\beta \xi$, so that $\beta$ may be interpreted as 
the fraction of the momentum on the pomeron carried by the scattered parton. 
Of course, these kinematics are valid not only for pomeron, but for {\em any} 
exchange.
\subsection{The diffractive structure function $F_2^{D(3)}$}
In an analogy with the structure function $F_2(x,Q^2)$, 
the 4-variable diffractive structure function (DSF) of the proton 
is defined through the  equation
\begin{equation}
\frac{d^4\sigma}{dQ^2\,d\beta\, d\xi\, dt}=
\frac{4\pi \alpha^2}{\beta\, Q^4}\left(2-2y+\frac{y^2}{2(1+R)}\right)
\cdot F_2^{D(4)}(Q^2,\beta,\xi,t)
\label{F2D4}
\end{equation}
The H1 and ZEUS data presented in this paper are integrated over $t$ 
within the region $|t_{min}|<|t|<1$ GeV$^2$. The $t$-integrated DSF,
$F_2^{D(3)}$, 
was determined from the data by assuming $R=0$ in (\ref{F2D4}).  Thus, 
the measured $F_2^{D(3)}$ 
is defined by
\begin{equation}
\frac{d^3\sigma}{dQ^2 \,d\beta \,d\xi}\,\stackrel{|t|<1}{\equiv}\,
\frac{4\pi \alpha^2}{\beta Q^4}\left(2-2y+\frac{y^2}{2}\right)
\cdot F_2^{D(3)}(Q^2,\beta,\xi)
\label{F2D3}
\end{equation}
\subsection{Gap factorization of $F_2^{D(3)}$}
From an analysis of their 1993 data, 
both the H1~\cite{H193} and ZEUS~\cite{ZEUS93} Collaborations found that 
$F_2^{D(3)}$ has the form
\begin{equation}
F_2^{D(3)}(Q^2,\beta,\xi)=\frac{1}{\xi^{n}}\cdot A(Q^2,\beta)
\label{F2D3-exp}
\end{equation}
The kinematic range of the measurements was

\begin{tabular}{lll}
H1&$3.0\times 10^{-4}<\xi<0.05$&$8.5<Q^2<50$ GeV$^2$\\
ZEUS&$6.3\times 10^{-4}<\xi<0.01$&~$10<Q^2<63$ GeV$^2$\\ 
\end{tabular}

The factorized form (\ref{F2D3-exp}) 
is what is expected from the IS model~\cite{IS}, in which the $F^{D(4)}_2$ 
is given by  the pomeron flux times the pomeron structure function:
\begin{equation}
F^{D(4)}_2(Q^2,\beta,\xi,t)
=f_{\pom /p}(\xi,t)\cdot F_2^{\pom}(Q^2,\beta)
=\frac{K}{\xi^{1+2(\epsilon+\alpha't)}}F^2(t)
\cdot F_2^{\pom}(Q^2,\beta)
\label{F2D4-IS}
\end{equation}
Since $\xi$ is related to the rapidity gap, we will refer to 
this form of factorization as {\em gap factorization}. 
The parameter $n$ of the $\xi$-dependence 
in (\ref{F2D3-exp}) corresponds to the $t$-averaged value of  
$1+2(\epsilon+\alpha't)$ in (\ref{F2D4-IS}), which we shall denote by 
$n_{soft}$ to indicate its connection to the pomeron intercept 
obtained from soft collisions (total cross sections). With $\epsilon=0.104$, 
$\alpha'=0.25$ GeV$^{-2}$,
and $<|t|>\approx 0.14$ GeV$^2$ (for the $\xi$-range 
of the experiments), we obtain $n_{soft}=1.14$. The values obtained by 
H1 and ZEUS are $n_{H1}=1.19\pm 0.06(stat)\pm 0.07(syst)$
and $n_{ZEUS}=1.30\pm 0.08(stat)^{+0.08}_{-0.14}(syst)$.
Below, in a separate subsection, we will discuss the significance of the 
pomeron intercept measured at high $Q^2$.
\subsection{Breakdown of gap factorization}
Rapidity gaps are expected to occur not only by the exchange of a pomeron,
but also by the exchange of a meson. The Regge trajectories of the light
mesons fall into three groups~\cite{CMG}: 
\begin{equation}
\alpha_{f/a}(t)=0.68+0.82\,t\;\;\;\;
\alpha_{\omega/\rho}(t)=0.46+0.92\,t\;\;\;\;
\alpha_{\pi}(t)=0+0.7\,t
\label{trajectories}
\end{equation}
Assuming that gap factorization holds for each exchange, 
$F_2^{D(4)}$ takes the form
\begin{equation}
F^{D(4)}_2(Q^2,\beta,\xi,t)
=\sum_{k} f_{k/p}(\xi,t)\cdot F_2^{k}(Q^2,\beta)
\label{F2D4M}
\end{equation}
where $k=\pom,\,f/a,\,\omega/\rho$ or $\pi$,  
$f_{k/p}(\xi,t)\sim \xi^{1-2\alpha_k(t)}$ 
is the flux and $F_2^{k}(Q^2,\beta)$ the structure 
function of the particle represented by $k$. 
Interference terms between trajectories could also be added in (\ref{F2D4M}).
Gap factorization will break down if the the pomeron $F_2$ structure 
is different than the meson structure. Motivated by (\ref{F2D4M}),
the H1 Collaboration fitted~\cite{H194}  their 1994 data with the form
\begin{equation}
F_2^{D(3)}(Q^2,\beta,\xi)=F_2^{\pom}(Q^2,\beta)\cdot \xi^{-n_{\pom}}+
C_M\cdot F_2^{M}(Q^2,\beta)\cdot \xi^{-n_{M}}
\label{F2D3M}
\end{equation}
using as $F_2^{M}(Q^2,\beta)$ the GRV parametrization for the pion 
structure function and treating all other parameters as free. The fit yielded 
$n_{\pom}=1.29\pm 0.03(stat)\pm 0.07(syst)$ and 
$n_{M}=0.3\pm 0.3(stat)\pm 0.6(syst)$. The trajectories calculated from these
values, after accounting for the t-dependence, are 
$\alpha_{\pom}=1.18\pm 0.02(stat)\pm 0.04(syst)$ and
$\alpha_{M}=0.6\pm 0.1(stat)\pm 0.3(syst)$. The value of $\alpha_{M}$ 
is consistent with the value of $\alpha_{f/a}(0)$ in (\ref{trajectories}).
\subsection{The pomeron intercept}
There have been theoretical speculations, based on the 
Balisky-Fadin-Kuraev-Lipatov (BFKL) QCD model of the pomeron~\cite{BFKL}, 
that the effective pomeron intercept, $\alpha_{\pom}(0)$, may increase 
with increasing $Q^2$. The pomeron participating at high $Q^2$ interactions 
is generally referred to as ``hard pomeron", to distinguish it 
from the ``soft pomeron" participating in soft processes. The question 
as to whether there exist two different pomerons is currently the subject of 
intense theoretical debate.  

Experimentally, there is no conclusive 
evidence for two different pomerons, despite trends in the data that may 
advocate the contrary. Where signs of a ``hard" pomeron appear in the 
data, usually other, more mundane, interpretations are possible. 
The pomeron intercept is a good example. Let us compare the values of the 
intercepts measured at high $Q^2$ at HERA with the value of the soft 
pomeron intercept obtained from the total cross sections:
\begin{table}[h]
\caption{The pomeron intercept $\alpha_{\pom}(0)=1+\epsilon$}
\begin{center}
\begin{tabular}{ll}
Measurement&Intercept\\
\hline
Soft~\cite{CMG}&$1.104\pm 0.002$\\
ZEUS-93&$1.18\pm 0.04^{+0.04}_{-0.07}=1.18^{+0.06}_{-0.08}$\\
H1-94&$ 1.18\pm 0.02\pm 0.04=1.18\pm 0.05$\\
ZEUS-94&$\epsilon_{hard}>\epsilon_{soft}$ (see text)\\
\end{tabular}
\end{center}
\label{intercept}
\end{table}

The HERA intercepts are systematically higher than the soft 
intercept, although statistically the effect is $\sim 1.5\sigma$. 
The question then is: do we really see the onset of the ``hard" pomeron?
Before answering this question, we comment on the ZEUS-94 measurement 
(last entry in Table~\ref{intercept}).  ZEUS measured the intercept from the 
cross section for $\gamma^*p\rightarrow Xp$ as a function of $W^2$
in the $Q^2$ range from 8 to 60 GeV$^2$.  For a fixed diffractive mass, 
the cross section should vary as $(W^2)^{2\epsilon}$.  
In two data sets, one with $M_X<3$ GeV and the other with $3<M_X<7.7$ GeV,
the measured intercept values are in 
agreement with the H1 value, but show signs of a systematic 
increase with $Q^2$ within the limited statistics of the available data.
Once again,  is this a sign of the onset of the ``hard" pomeron?

There are two important aspects of 
the ZEUS-94 data samples, which are evident from the relations 
$\beta\approx Q^2/(Q^2+M_X^2)$ and 
$\xi\approx (Q^2+M_X^2)/W^2$ obtained from (\ref{DISD}):
\begin{itemize}
\item $\beta$ is high, so that there is no substantial 
meson-exchange contribution. 
\item For fixed $M_X^2$ and $Q^2$, $W^2\sim \xi^{-1}$. Therefore, 
the $W^2$-dependence of the cross section is, in effect, a $\xi$-dependence. 
Moreover, when $M_X^2$ is small and kept fixed, {\em regions of 
larger values of $\xi$ 
are probed as \mbox{$Q^2$} increases}!
\end{itemize}
Thus, what is perceived as an increase of the value of the intercept with 
$Q^2$, is in fact a steepening of the $\xi$-distribution with increasing $\xi$.
Such a steepening may be due to the de-coherence expected due to hadronic 
final state 
interactions, which should increase as the rapidity gap
decreases. 

The steepening of the $\xi$-distribution should be more prominent in the 
Tevatron high $E_T$ diffractive dijet data, and in particular in the 
CDF ``Roman Pot" data, for which $0.05<\xi<0.1$. Such an effect 
would reduce the dijet production rate at high $\xi$ relative to that 
at small $\xi$. 

In conclusion, the apparent increase of the pomeron intercept with $Q^2$ 
may be due to a gradual steepening of the $\xi$-distribution with 
increasing $\xi$ caused by a de-coherence effect 
due to final state interactions 
as the rapidity gap decreases. A measurement of the diffractive dijet rate 
as a function of $\xi$ at the Tevatron 
can provide a decisive test of this hypothesis. 

\subsection{The structure of the pomeron}

The shape ($Q^2$ and $\beta$ dependence) of the pomeron structure function 
$F_2^{\pom}(Q^2,\beta)$, presumed to be the term in $F_2^{D(3)}$ that 
contains the $(Q^2,\beta)$-dependence,  was obtained by H1~\cite{H193} and 
ZEUS~\cite{ZEUS93}  using fits of the form (\ref{F2D3-exp}) on their 
1993 data, and by H1\cite{H194} using fit (\ref{F2D3M}) on their 1994 
data. All fits yielded an approximately 
flat $\beta$-distribution, within the measured 
range of $0.01<\beta<0.9$, and a $Q^2$-distribution which, for $\beta<0.65$, 
increases slowly ($\sim$logarithmically) by a factor of 
$\sim 1.5$ between $Q^2=10$ 
and $Q^2=60$ GeV$^2$. Above $\beta=0.65$, the data (H1), which are all  
in a single  $\beta$-bin centered  at $\beta=0.9$, show a flat 
$Q^2$ dependence. However, this $\beta$-bin includes a substantial 
contribution 
from vector mesons, whose relative magnitude decreases as $Q^2$ increases.

The gluon content of the pomeron was determined  by ZEUS~\cite{ZEUSJETS} 
and H1~\cite{H194} using different techniques. ZEUS determined it
by a combined analysis of the rate of diffractive DIS, which is 
mainly sensitive to the quark content of the pomeron, and the rate of 
diffractive 
inclusive jet production, $ep\rightarrow jet+X$ ($E_T^{jet}>8$ GeV), which is 
sensitive to both the quark and gluon contents. This analysis yielded 
a hard-gluon fraction of $0.3<f_g<0.8$, {\em i.e.} 
30\% to 80\% of the momentum of the 
pomeron carried by its partons is due to hard gluons. 
H1 derived the gluon content 
from the $Q^2$-dependence of the $F_2^{D(3)}$ structure function.
By interpreting this dependence as arising from scaling violations, and 
fitting the data using the 
DGLAP evolution equations, a QCD analysis led to the conclusion 
that gluons carry about 80\% of the momentum of the pomeron at 
$Q^2\sim Q_{\circ}^2=2.5$ GeV$^2$, and are concentrated at $\beta_{g/\pom}=1$.
As $Q^2$ increases, the H1 analysis shows that the gluon content decreases 
slowly, while the quark content increases. However, these variations are 
very slow, so that at $Q^2$ values from $\sim 25$ to $\sim 1000$ GeV$^2$ the 
gluon fraction decreases from $\sim 80\%$ to $\sim 70\%$, and the 
quark fraction increases from $\sim 20\%$ to $\sim 30\%$, with both 
distributions remaining fairly hard (this conclusion is significant in 
using the H1 structure function of the pomeron to predict 
diffractive $W$ and dijet production rates at the Tevatron).

Despite its appeal, the interpretation of the $Q^2$-dependence of 
$F_2^{D(3)}$ as being due to scaling violatios
is not unique. Goulianos argued~\cite{R} that 
the observed $Q^2$-dependence 
is due to the pomeron flux renormalization, which 
we referred to earlier as {\em scaled gap probability}. 
A good fit to the data was obtained and, based on this fit, 
predictions for the rates of $W$ and dijet production at the Tevatron were made.
As will be shown below, in the section ``HERA versus Tevatron", 
the renormalized/scaled flux predictions 
have now been confirmed by the CDF and D\O\, data.
\section{Results from the Tevatron}
Results on hard diffraction from the Tevatron have been reported by both the
CDF and D\O\,Collaborations. One set of results was obtained using 
rapidity gaps as a tag for diffraction, while another by detecting  
the leading antiproton in a Roman Pot spectrometer.
Figure~4 shows the event topology for dijet production in 
single diffraction, double diffraction
and double pomeron exchange. Due to space limitations, double diffraction 
will not be discussed further. 
\vglue -0.15in
{\hspace*{-0.75in}\psfig{figure=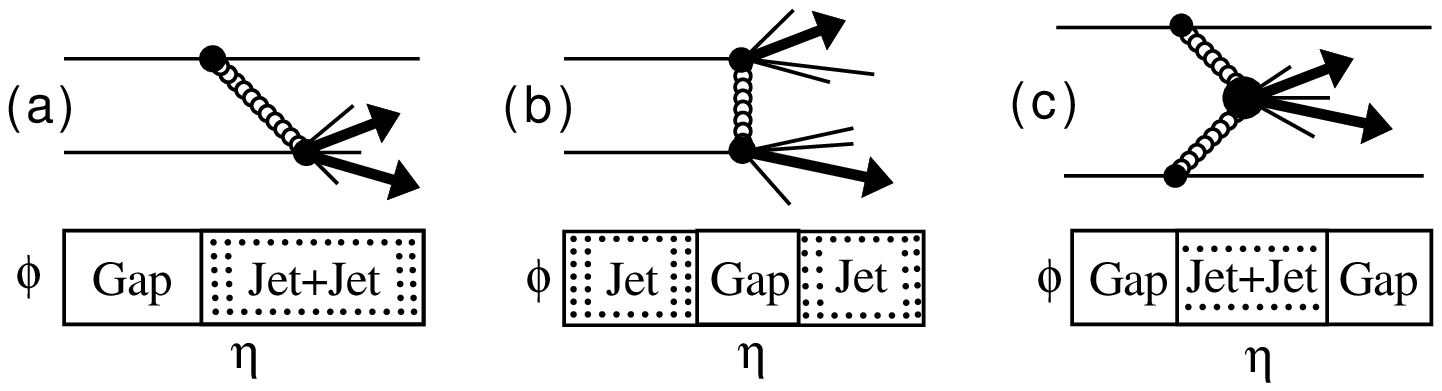,width=6.2in}}
\vglue -6.75in
\centerline{Figure 4: Dijet production diagrams and event topologies for}
\centerline{(a) single diffraction (b) double diffraction
and (c) double pomeron exchange.}
\vglue 0.1in
Below, we will discuss first the rapidity gap results and then some
Roman Pot results. All results are for $p\bar p$ collisions at 1.8 TeV 
unless otherwise stated. A confrontation between these results 
and predictions based on the diffractive 
structure function measured at HERA is planned for the next section.
\subsection{Diffractive $W$ production (CDF)}
CDF made the first observation~\cite{W}  
of diffractive $W$ production and measured the production 
rate using a sample of 8246 events with an isolated 
central $e^+$ or $e^-$ ($|\eta|<1.1$)  of $E_T>20$ GeV 
and missing transverse energy $\not\!\! E_T>20$ GeV. In searching for
diffractive events, CDF studied the correlations of 
the Beam-Beam Counter (BBC) ($3.2|\eta|<5.9$) multiplicity, $N_{BBC}$, 
with the sign of the electron-$\eta$, $\eta_e$,  or  
the sign of its charge, $C_e$. 
In a diffractive $W^{\pm}\rightarrow e^{\pm}\nu$ event produced in
a $\bar p$ collision with a pomeron emitted by the proton,
a rapidity gap is expected at positive $\eta$ ($p$-direction),
while the lepton is boosted towards negative $\eta$ (angle-gap correlation).
Also, since the pomeron is quark-flavor symmetric,
and since, from energy considerations, mainly valence quarks from the
$\bar p$ participate in producing the $W$,
approximately twice as many electrons as positrons
are expected (charge-gap correlation). 

\noindent\begin{minipage}[t]{7cm}
.
\vglue -1.8cm
{\hspace*{-0.5cm}\psfig{figure=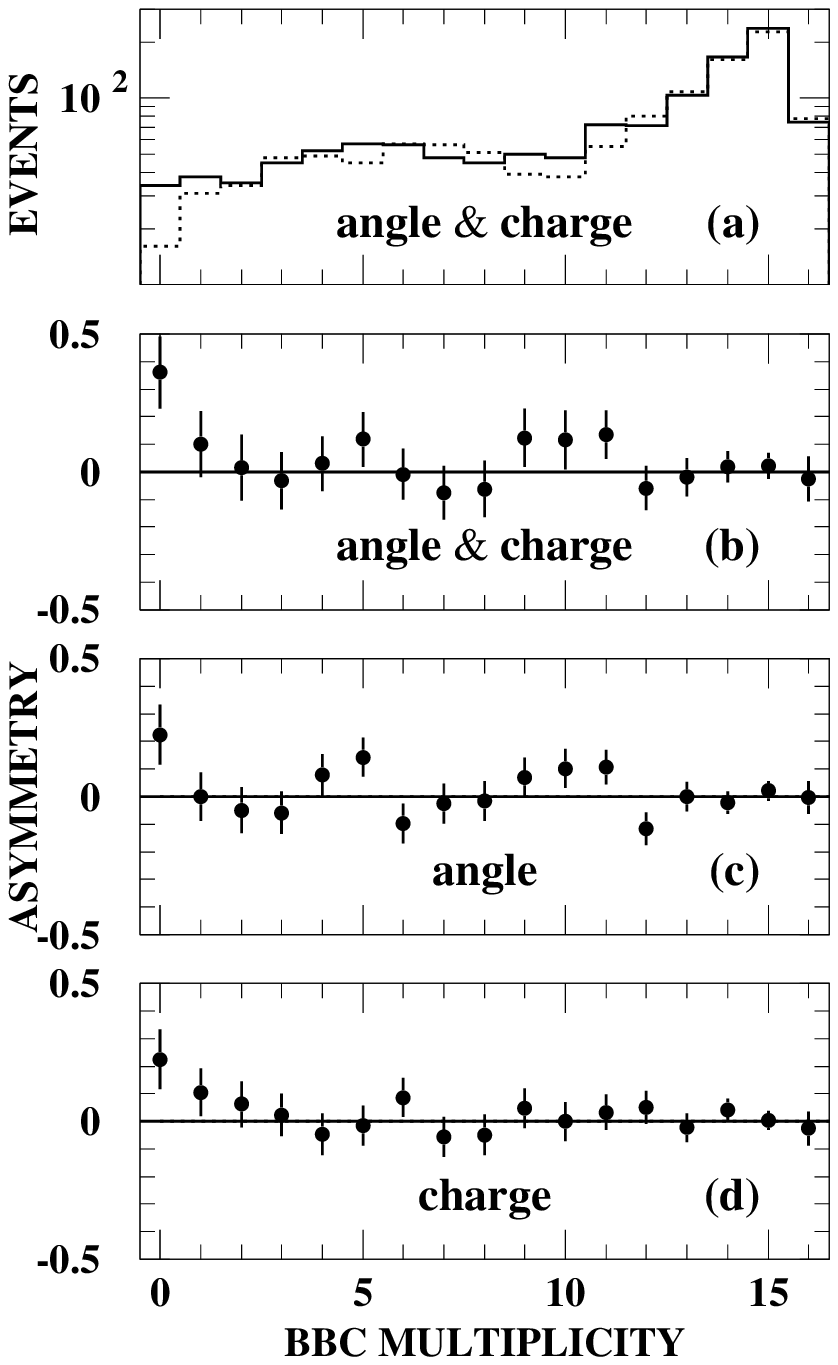,width=15cm}}
\vglue -3cm
\centerline{Figure 5: The diffractive $W$ signal.}
\end{minipage}
\noindent\begin{minipage}[t]{4.8cm}
Figure~5 shows $N_{BBC}$ distributions for two event samples, 
one with $\eta_e\,\eta_{BBC}<0$ (angle-correlated)
or $C_e\, \eta_{BBC}<0$ (charge-correlated), and the other for 
$\eta_e\, \eta_{BBC}>0$ (angle-anticorrelated) or
$C_e\, \eta_{BBC}>0$ (charge-anticorrelated).
The {\em doubly-correlated (anticorrelated}, dotted)
distributions, shown in Fig.~5a,
are for events with $\eta_e\,  C_e>0$ and $\eta_e\, \eta_{BBC}<0$
($\eta_e \, \eta_{BBC}>0$).  Figure~5b 
shows the bin-by-bin asymmetry (difference divided by sum)
of the two distributions of Fig.~5a.
The excess seen in the first bin is the 
signature expected from diffractive events with a
rapidity gap.  An excess is also seen
in the individual angle (Fig.~5c) and charge (Fig.~5d)
asymmetries, as expected.
The probability that the observed excess is caused by
fluctuations in the non-diffractive
background was estimated to be $1.1\times 10^{-4}$. 
\end{minipage}

\noindent Correcting for acceptance,
the ratio of diffractive to non-diffractive $W$ production is found to be:\\
\centerline{$R_W=[1.15\pm 0.51(stat)\pm 0.20(syst)]\%\;\,(\xi<0.1)$}\\
The standard flux prediction for a two (three) flavor full hard-quark
pomeron structure is $R_W^{hq}$=24\% (16\%) and for a full hard-gluon
structure $R_W^{hg}=1.1\%$. 
The measured ratio favors
a purely gluonic pomeron, but this is incompatible with
the low fraction of diffractive $W+Jet$ events observed. 
A more complete comparison with theoretical 
predictions is made below in combination with 
the diffractive dijet CDF result.
\subsection{Diffractive dijet production (CDF)}
CDF searched for diffractive dijet production in a sample of 30352 dijet 
events with a single vertex 
(to exclude events from multiple interactions), in which  
the two leading jets 
have $E_T>20$ GeV and are both at $\eta<1.8$ or $\eta>1.8$.
No requirement was imposed on the presence or kinematics of extra jets in 
an event. Figure~6 shows the correlation of the BBC and forward 
($|\eta|>2.4$) calorimeter 
tower multiplicities in the $\eta$-region opposite the dijet system. The 
excess in the 0-0 bin is attributed to diffractive production. 
After subtracting the non-diffractive background and 
correcting for the single-vertex selection cut, for 
detector live-time acceptance  and for the 
rapidity gap acceptance $(0.70\pm 0.03)$, calculated using the POMPYT 
Monte Carlo program \cite{W} with pomeron $\xi<0.1$, the 
``Gap-Jet-Jet" fraction (ratio of diffractive to non-diffractive dijet events)
was found to be 
\begin{center}
$R_{GJJ}=[0.75\pm 0.05(stat)\pm 0.09 (syst)]\%=(0.75\pm 0.10)\%$\\ 
($E_T^{jet}>20$ GeV, $|\eta|^{jet}>1.8$, $\eta_1\eta_2>0$, $\xi<0.1$)
\end{center}
\vglue -1cm
\noindent\begin{minipage}[t]{5cm}
{\psfig{figure=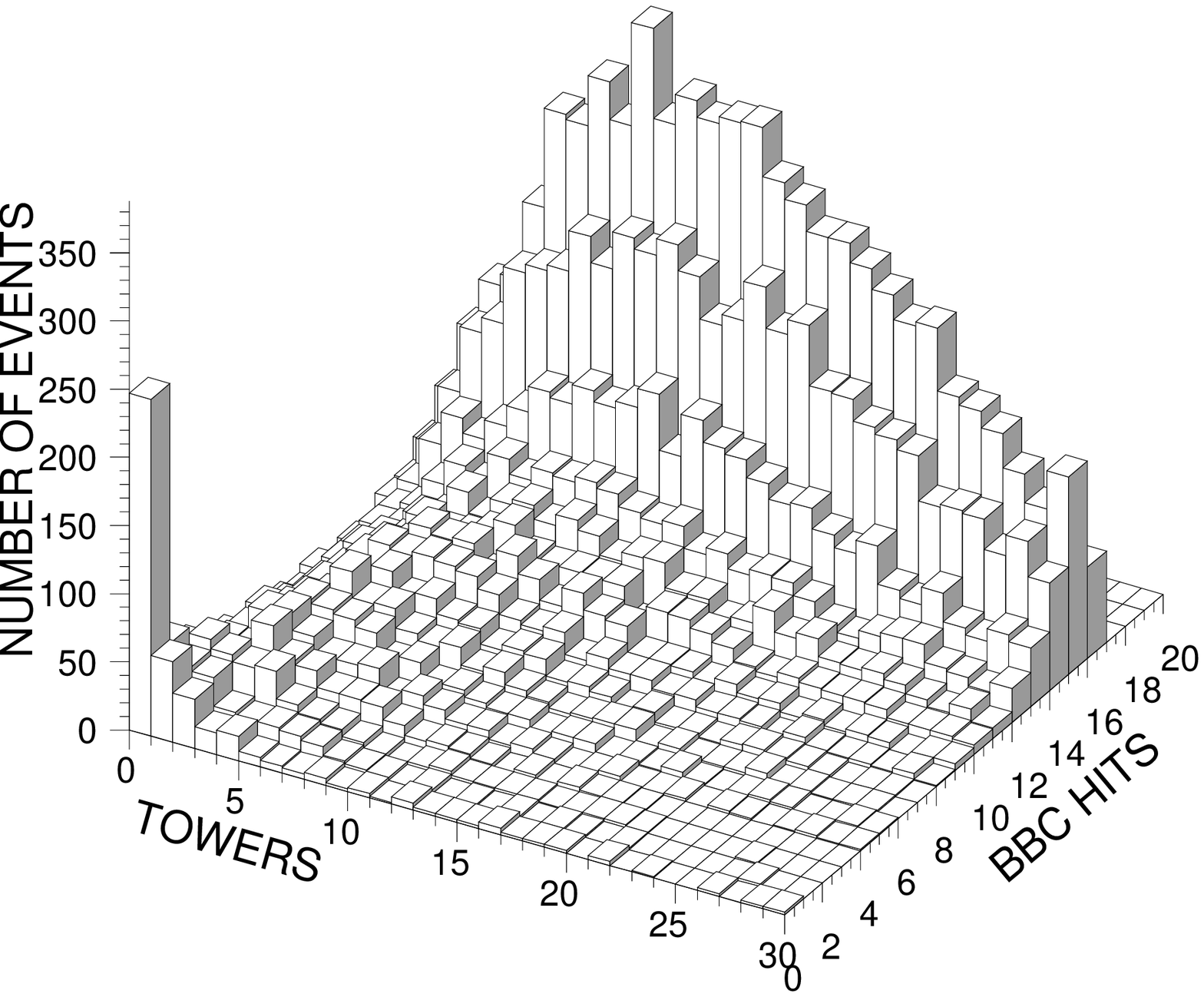,width=2.75in}}
{\small 
Figure 6: Tower versus BBC multiplicity for dijet events with both jets at
$\eta>1.8$ or $\eta<1.8$.}
\end{minipage}
\hspace*{-0.5cm}
\noindent\begin{minipage}[t]{5.5cm}
\psfig{figure=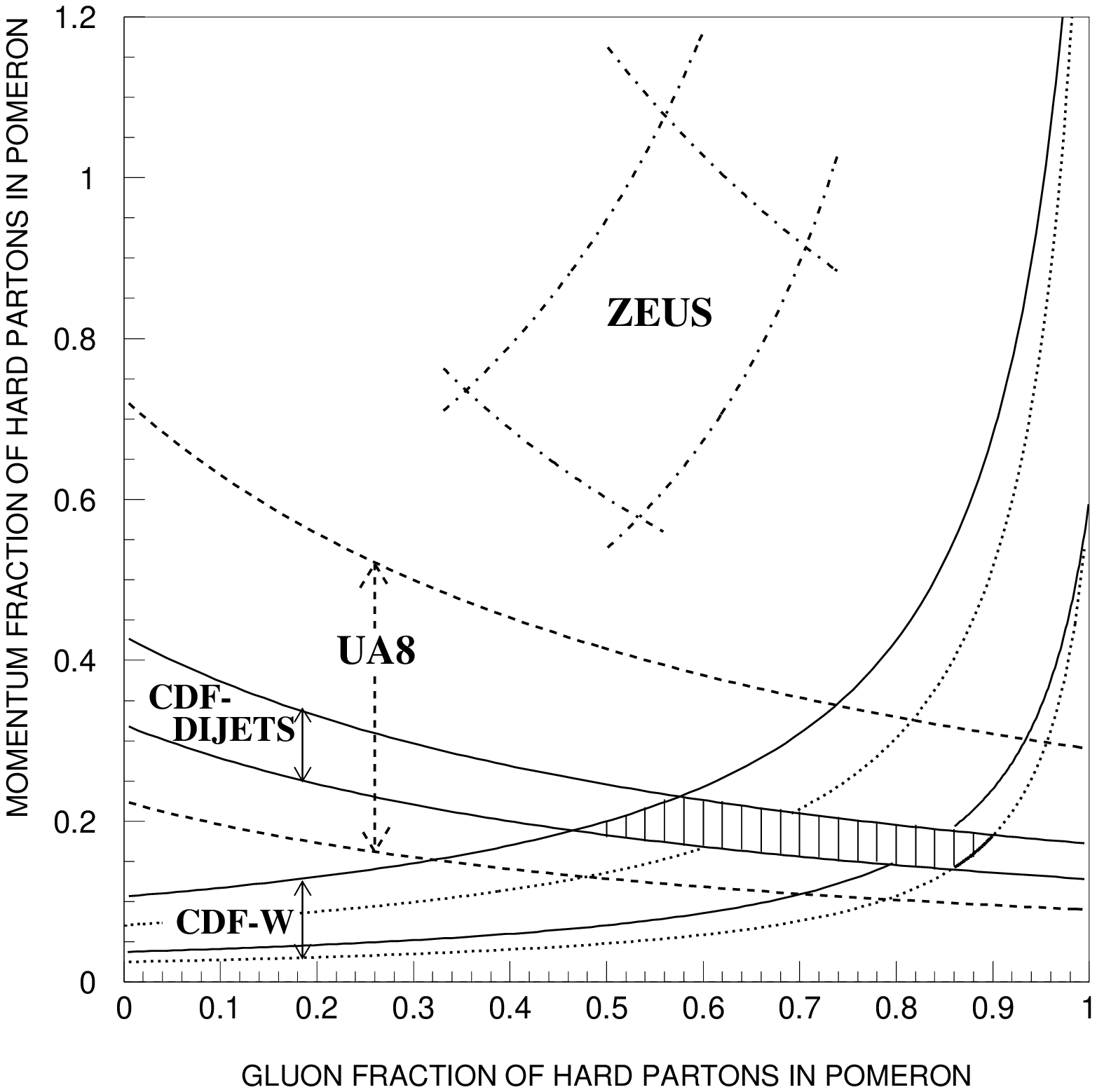,width=2.32in}
{\small 
Figure 7: Total momentum fraction of partons in the pomeron versus 
gluon fraction.}
\end{minipage}
\vglue 0.5cm
\subsection{The gluon fraction of the pomeron (CDF)}
By combining the diffractive $W$ and dijet results, CDF extracted 
the gluon fraction of the pomeron, $f_g$. 
Assuming the standard pomeron flux, the measured $W$ and dijet fractions  
trace  curves in the plane of $D$ versus $f_g$, where $D$ is the total momentum 
fraction carried by the quarks and gluons in the pomeron. Figure~7 shows the 
$\pm1\sigma$ curves corresponding to the results. 
From the diamond-shaped overlap of the $W$ and dijet 
curves,  CDF obtain $f_g=0.7\pm 0.2$. This result, which is independent of the 
pomeron flux normalization, agrees with the result obtained by 
ZEUS~\cite{ZEUSJETS}
from DIS and dijet photoproduction (dashed-dotted line in Fig.~7).
For the $D$-fraction, CDF obtain the value 
$D=0.18\pm 0.04$. 
In the next section  we will show that the decrease of the 
$D$-fraction at the Tevatron relative to that at HERA
can be accounted for by the pomeron flux renormalization factor. 
The dashed lines are the 
$\pm 1\sigma$ curves for the UA8 diffractive dijet data.
To compare UA8 with CDF, the UA8 fractions must first be multiplied by the 
ratio of the renormalization factors at the two energies, which  is~\cite{R}
$D_{CDF}/D_{UA8}\approx 0.7$. Within the errors, the results of the two
experiments are in good agreement.
\subsection{Diffractive dijet production (D\O\,)}
D\O\, measured diffractive dijet production
at $\sqrt{s}=$1800 and 630 GeV using a rapidity gap tag. 
The D\O\, results, which are not corrected for 
acceptance, are shown in Fig.~8. The acceptance is expected 
to be of ${\cal{O}}(70\%)$. Thus, the D\O\, 1800 GeV fraction is in general 
agreement with the CDF measurement.\\
\vglue 0.25cm
\hbox{\epsfysize=6cm\epsfxsize=6cm\epsffile{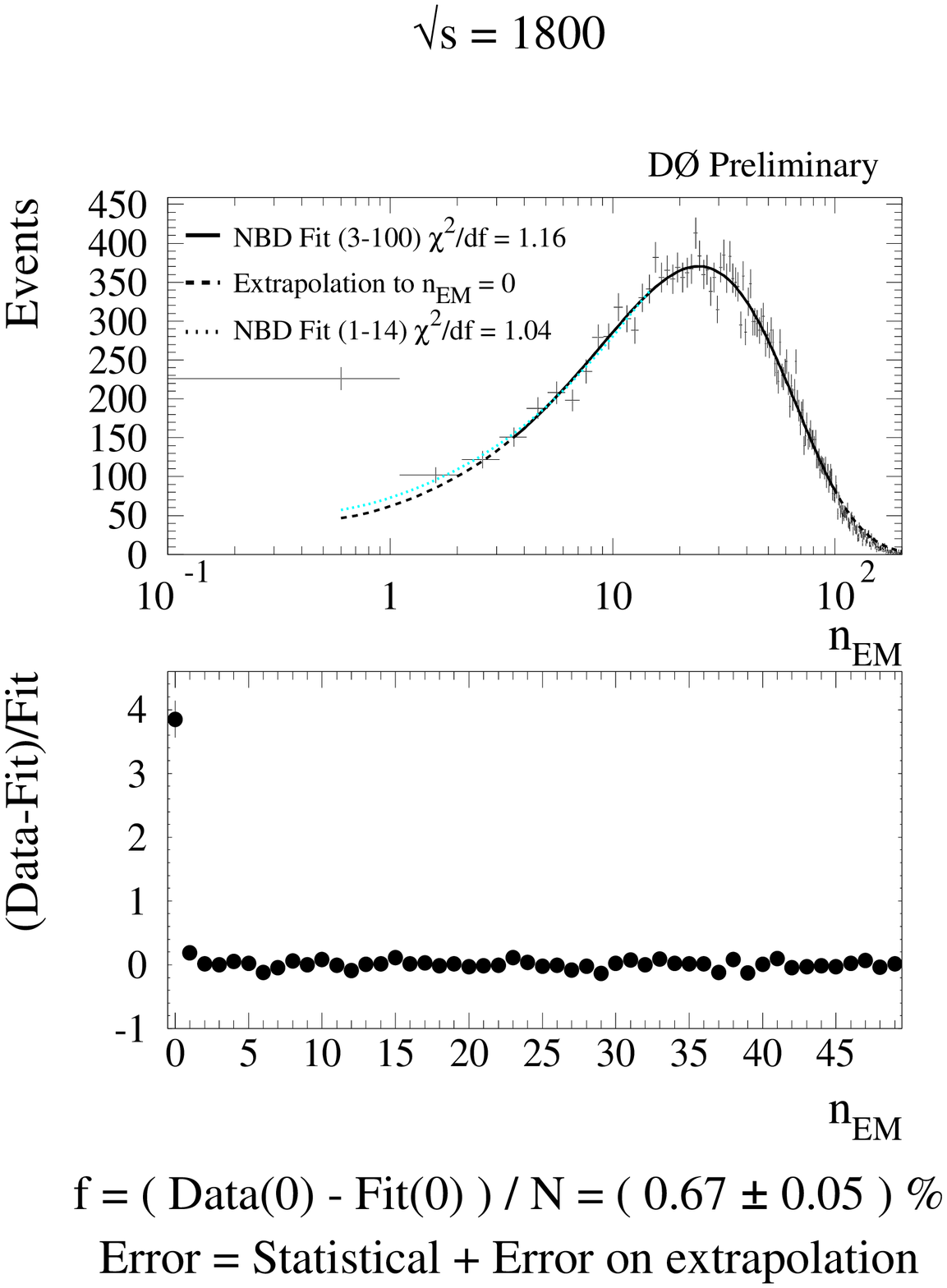}\hspace*{0.5cm}\epsfysize=6cm\epsfxsize=6cm\epsffile{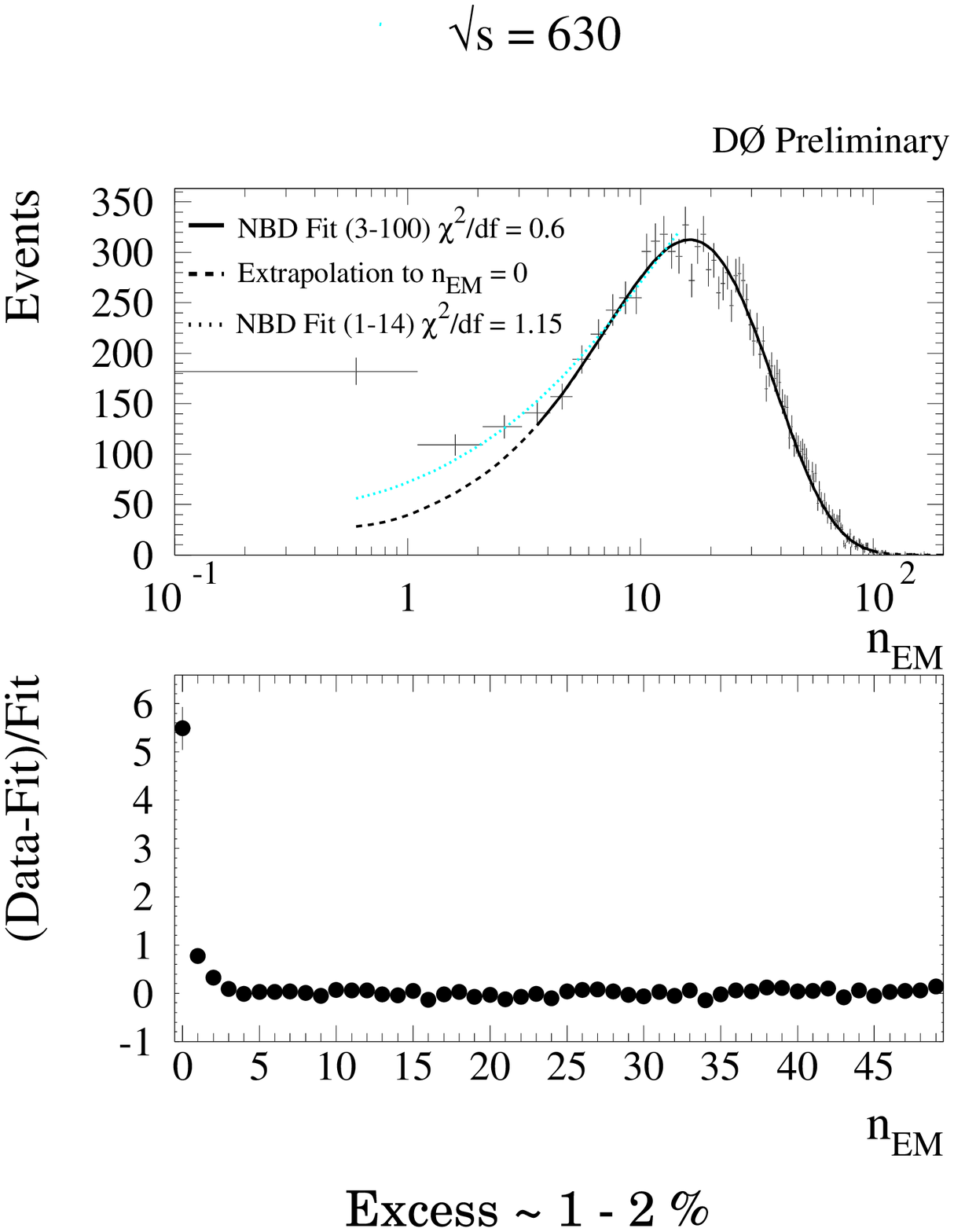}}

\vglue 0.15in
\begin{center}
Figure~8: D\O\, results on diffractive dijet production:\\
{\em (left)} at $\sqrt{s}=1800$ GeV and {\em (right)} at $\sqrt{s}=$630 GeV.\\
The top figures show the calorimeter tower multiplicity in the forward
region ($|\eta|>2$) opposite the jets, and a fit to the data; the bottom 
figures show the ratio (data-fit)/fit. 
The excess in the first bin is the diffractive dijet 
signal.
\end{center}

\subsection{Dijets in double-pomeron exchange}
Both the CDF and the D\O\, Collaborations have observed dijet events
produced by the interaction of two pomerons, one emitted by the $p$ and the
other by the $\bar p$. Using rapidity gaps at $|\eta|>2$ to tag such events
(see Fig.~4), D\O\, measured the ratio of double-pomeron exchange (DPE) to
non-diffractive (ND) dijet events to be
\begin{center}
$\mbox{D\O\, result: }\sigma
\left(\frac{DPE}{ND}\right)_{E_T^{jet}>15\;GeV}={\cal{O}}(10^{-6})$
\end{center}
CDF tagged DPE dijet events by observing the leading antiproton in a Roman Pot
spectrometer (Fig.~9) and a rapidity gap on the side of the proton.
\vglue -0.25in
\centerline{\psfig{figure=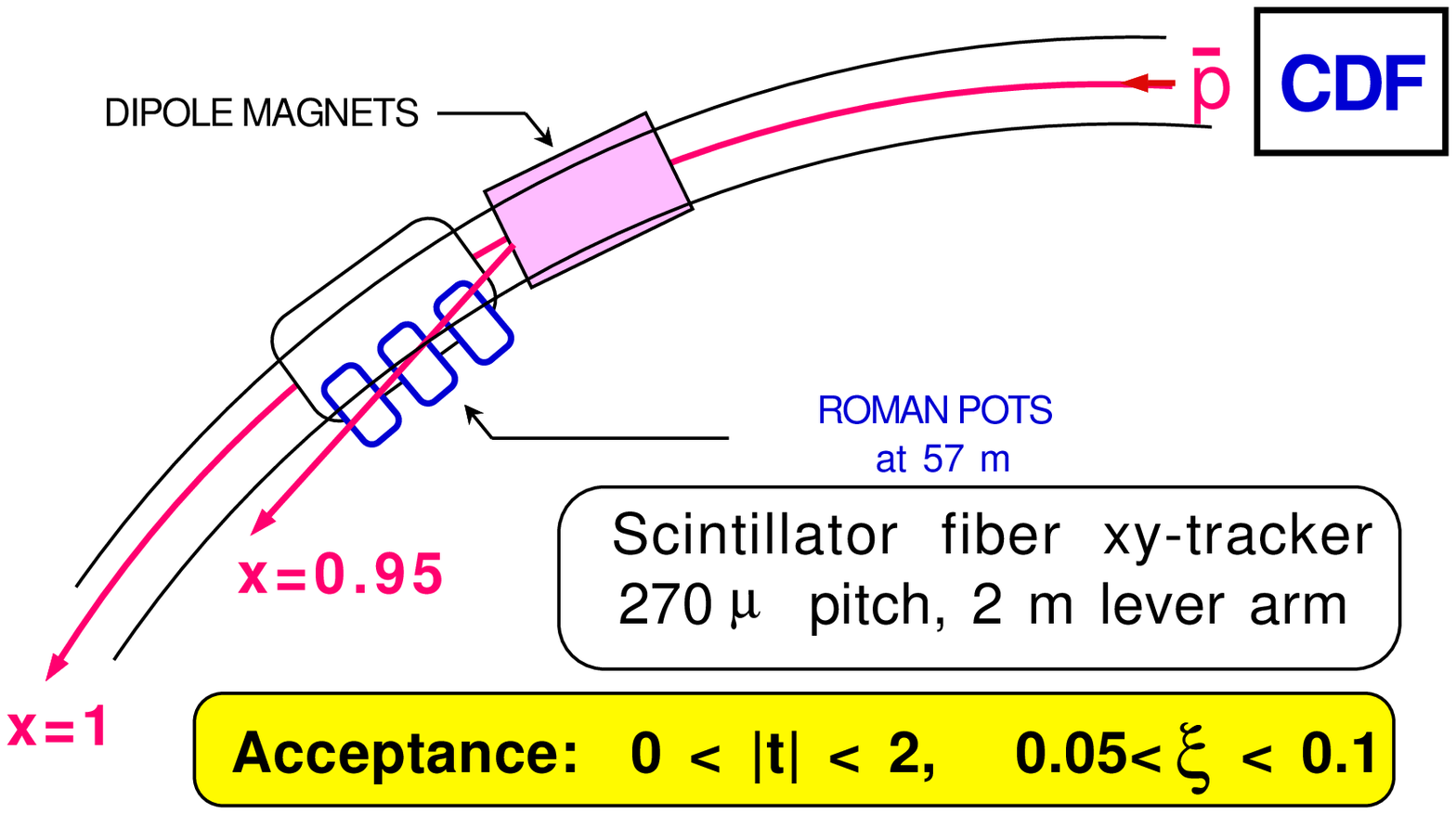,width=5in}}
\vglue -3.8in
\centerline{Figure 9: Schematic view of the CDF leading antiproton spectrometer}
\vglue 0.1in
The $\bar p$ trigger required the $\xi$ of
the pomeron from the $\bar p$ to be within
the range $0.05<\xi_{\pom/\bar p}<0.1$,
while from the rapidity interval covered by
the BBC and from energy considerations it is estimated that the
$\xi$ of the pomeron on the $p$ side lies approximately within the range
$0.015<\xi_{\pom/p}<0.035$. With these $\xi$-values, the energy in the
$\pom-\pom $ center of mass system,
$(\xi_{\pom/\bar p}\cdot \xi_{\pom/_p}\cdot s)^{\frac{1}{2}}$, is
approximately in the range 50-100 GeV.
The DPE cross section was compared with the SD and 
non-diffractive (ND) dijet cross sections for $E_T^{jet}>7$ GeV:
\begin{center}
\vglue 0.1in
\begin{tabular}{ll}
DPE/SD&$[0.170\pm 0.036(stat)\pm 0.024(syst)]\%$\\
SD/ND&$[0.160\pm 0.002(stat)\pm 0.024(syst)]\%$\\
(DPE/SD)/(SD/ND)&\fbox{$1.1\pm 0.3$}\\
DPE/ND&$(2.7\pm 0.7)\times 10^{-6}$\\
\end{tabular}
\end{center}
Assuming that both the $p$ and $\bar p$ pomeron fluxes are renormalized, 
the ratio enclosed in the box is expected to be $\approx 1$. 
\section{HERA versus Tevatron}
The diffractive structure function measured in DIS at HERA can be used 
directly in PYTHIA to calculate the rate of diffractive $W$-boson production 
in $p\bar p$ collisions at the Tevatron. Such a calculation 
yields~\cite{KGDIS} $R_W=6.7\%$ for the ratio of diffractive to non-diffractive 
$W$ production
(a similar result has been 
obtained~\cite{Whitmore} by L. Alvero, J. Collins, J. Terron and J. Whitmore).
The experimental value\cite{W} of 
$(1.15\pm 0.55)$\% is smaller than 6.7\% by a factor of $0.17\pm 0.08$. 
The deviation  of this factor from unity represents a 
breakdown of {\em QCD factorization}. 

It has been shown~\cite{KGDIS} that, assuming that the 
gap probability distribution in the  DSF of (\ref{F2D3}) 
scales to the total gap probability, the 
ratio of the scaling factors between Tevatron and HERA is 
$D_{scale}=0.19$.  This factor agrees with the value of $0.17\pm 0.08$ 
and with the momentum fraction of 
$0.18\pm 0.04$ measured by CDF from the $W$ and dijet rates.
As mentioned earlier, the scaling of the gap probability 
is, in effect, the pomeron flux renormalization scheme proposed~\cite{R} to 
unitarize the soft diffraction cross section. Thus, the breakdown of
QCD factorization in hard diffraction is traced back to the breakdown of 
factorization in soft diffraction, which in itself is dictated by 
unitarity and is expressed as a scaling law of the gap probability 
distribution.

\section{Conclusion}
Experiments at HERA and at $p\bar p$ Colliders show that the pomeron
has a hard partonic structure, which is dominated by gluons but 
has a substantial quark component. The DSF 
of the proton, measured in DIS at HERA, 
fails to predict the diffractive $W$ and dijet 
rates measured at the Tevatron.  This breakdown of factorization is restored 
by scaling the rapidity gap probability distribution, which appears as 
a factor in the DSF,
to the total gap probability.
The scaling of the 
gap probability is also needed to explain the s-dependence of the 
soft $p\bar p$  SD cross section. Assuming that the gap is caused 
by pomeron exchange, no breakdown of factorization occurs in the hard collision 
between the pomeron and the $\gamma^*$ at HERA or between the pomeron and 
the $p(\bar p)$ at the Tevatron.

Two questions that have been raised are: (a) Is the $Q^2$ 
dependence of the DSF due to QCD scaling violations or due to 
the scaling of the gap probability?
(b) Does the apparent increase of the pomeron intercept with $Q^2$ 
come from an increase with pomeron-$\xi$ due to hadronic final state 
interactions? 


\section*{References}
\begin{thebibliography}{99}
\bibitem{KG}K. Goulianos, Physics Reports {\bf 101} (1983) 169.
\bibitem{CMG} R.J.M. Covolan, J. Montanha and K. Goulianos,
Phys. Lett. {\bf B 389} (1996) 176.
\bibitem{IS} G. Ingelman and P. Schlein, Phys. Lett. {\bf B 152} (1985) 256.
\bibitem{POMPYT} P. Bruni and G. Ingelman,
Preprint DESY-93-187; Proceedings of the International Europhysics Conference
on High Energy Physics, Marseille, France, 22-28 July 1993, Editions
Fronti\`{e}res (Eds. J. Carr and M. Perrottet) p.595.
\bibitem{PYTHIA} T. Sj\"{o}strand, Comput. Phys. Comm. {\bf 82}, 74 (1994).
\bibitem{DLF} A. Donnachie and P. V. Landshoff, 
Nucl. Phys. {\bf B 303} (1998) 634.
\bibitem{R} K. Goulianos, Phys. Lett. {\bf B 358} (1995) 379; 
{\em ib.} {\bf B363} (1995) 268.
\bibitem{KGSX} K. Goulianos, Proceedings of the 3rd Workshop on Small-x and 
Diffractive Physics, Argonne National Laboratory, USA, 26-29 September 1996.
\bibitem{KGDIS} K. Goulianos, 
Proceedings of the 5th International Workshop 
on Deep Inelastic Scattering and QCD (DIS-97), Chicago, USA, 14-18 April 1997.
\bibitem{pbp2} K. Goulianos, Proceedings of the 10$^{th}$
Topical Workshop on Proton-Antiproton Collider Physics,
Fermilab, Batavia, IL, USA, 9-13 May 1995.
\bibitem{UA8} A. Brandt et al. (UA8 Collaboration),
Phys. Lett. {\bf B297} (1992) 417;\\
R. Bonino {\em et al.} (UA8 Collaboration), Phys. Lett. {\bf B211} (1988) 239.
\bibitem{UA8-EPS}P. Schlein, {\em Evidence for Partonic Behavior of the
Pomeron}, Proceedings of the International Europhysics Conference
on High Energy Physics, Marseille, France, 22-28 July 1993 (Editions
Frontieres, Eds. J. Carr and M. Perrottet).
\bibitem{H193} T. Ahmed {\em et al.},
Phys. Lett. {\bf B 348} (1995) 681.
\bibitem{ZEUS93}M. Derrick {\em et al.},
Z. Phys. {\bf C68} (1995) 569
\bibitem{H194}
H1 Collaboration: {\em A Measurement and QCD Analysis of the
Diffractive Structure Function} $F_2^{D(3)}$,
Contribution to ICHEP'96, Warsaw, Poland, July 1996.
\bibitem{BFKL}L. N. Lipatov, Sov. J. Nucl. Phys. {\bf 23}, 338 (1976);
E. A. Kuraev, L. N. Lipatov and V. S. Fadin, 
Sov. Phys. JETP {\bf 44}, 443 (1976);
Sov. Phys. JETP {\bf 45}, 199 (1977); Ya. Ya. Balisky and L. N. Lipatov,
Sov. J. Nucl. Phys. {\bf 28}, 822 (1978).
\bibitem{ZEUSJETS}M. Derrick {\em et al.},
Phys. Lett. {\bf B 356} (1995)  129.
\bibitem{W}F. Abe {\em et al.}, Phys. Rev. Lett. {\bf 78} (1997) 2698.
\bibitem{ssjet} F. Abe {\em et al.},
``Measurement of Diffractive Dijet Production at the Fermilab Tevatron",
submitted to Phys. Rev. Letters.
\bibitem{Whitmore}J. Whitmore,
Proceedings of the 5th International Workshop 
on Deep Inelastic Scattering and QCD (DIS-97), Chicago, USA, 14-18 April 1997.
\end{thebibliography}
\end{document}